# Deep-ultraviolet second harmonic generation in a conventional crystal quartz by freeing phase-matching in nonlinear optics

Mingchuan Shao[1], Fei Liang[2], Haohai Yu*, Huaijin Zhang**

State Key Laboratory of Crystal Materials and Institute of Crystal Materials, Shandong University, Jinan 250100, China.

*Corresponding author. E-mail: haohaiyu@sdu.edu.cn

**Corresponding author. E-mail: huaijinzhang@sdu.edu.cn

**Abstract:** Nonlinear frequency conversion in optics can originate the coherent light at the wavelength where it is hard or unlikely to achieve by directly lasing and is a fundamental topic in science and engineering covering both classical and quantum regions. The critical requirement for efficient nonlinear frequency conversion is phase-matching, a momentum conservation relation between the fundamental and harmonic light. It is dreaming of a technique for compensating the phase-mismatching in universal nonlinear materials and wavelength ranges, since the phase-matching was proposed in 1962. Here, an additional periodic phase (APP) concept was proposed for the phase-matching in nonlinear optics and experimentally demonstrated the APP phase-matched second harmonic generation with a conventional crystal quartz at the wavelength deep to vacuum-ultraviolet 177.3 nm. This study may not only develop a universal way to resuscitate the nonlinear optical materials for efficient nonlinear frequency conversion, but also may revolutionize the nonlinear photonics and their further applications.

Nonlinear optical process was first discovered in a crystal quartz by the realization of the second-harmonic generation (SHG) in 1961[1]. Not before long, the phase-matching, including quasi-phase-matching (a momentum relation between the fundamental and harmonic light), was proposed to provide constructive combing and interference of the microscopic nonlinear sources in the nonlinear medium and address the oscillation of harmonic fields along the propagation direction for achieving the high-efficiency frequency conversion[2,3]. Nowadays, associated with the natural birefringence of the nonlinear crystals at typical wavelength and periodic and/or aperiodic poling of the ferroelectric domains in the certain nonlinear photonic crystals[4-11], the coherent light in the wavelength region from the ultraviolet (UV) to terahertz has been developed and the nonlinear frequency conversion process has become a fundamental topic in numerous scientific and engineering applications that cover both classical and quantum regions[12-16]. However, the birefringence phase-matching and quasi-phase-matching conditions excluded the nonlinear materials that have no suitable birefringence or ferroelectric domains with reversible polarization sensitive to the external field[4,17]. Therefore, a long-standing goal in nonlinear optics is the discovery of a universal technique for realizing the phase-matching in arbitrary nonlinear media and wavelength ranges, since the phase-matching concept was proposed in 1962[2,3,18]. In the coherent light, the DUV is a fundamental source for the modern spectroscopic measurement, photolithographic techniques, material processing, etc.(e.g. to study the surfaces of superconductors and observe the band structure of topological materials[19-22]) and the generation of DUV by nonlinear frequency conversion is considered as the Holy Grail in nonlinear crystal optics[19]. Here, we proposed a universal solution for the phase-matching condition beyond those with the natural birefringence of the nonlinear crystals or artificially poling of the ferroelectric domains, and the efficient SHG output in the DUV region was experimentally demonstrated with a conventional crystal quartz.

Take the typical collinear frequency-doubling as an example. The electric field $E_{2\omega}(z)$ of SHG light is described as[4]

$$\frac{dE_{2\omega}(z)}{dz} = \frac{i\omega}{cn_{2\omega}} d_{eff}(z) E_\omega^2(z) e^{-i\Delta\varphi} \qquad (1)$$

where $E_\omega(z)$ denotes the electric field of the fundamental field at the propagation length $z$; $\omega$ refers to the fundamental frequency; c represents the light velocity; $n_{2\omega}(z)$ and $d_{eff}(z)$ denote the refractive indexes of the SHG light and effective nonlinear coefficient at the propagation length $z$, respectively; $\Delta\varphi = k_2 z - 2k_1 z = \Delta k z$ is the phase difference between the fundamental and SHG light with the wave vectors of $k_1$ and $k_2$. **Figure 1a** shows the schematic demonstration of the SHG performance of phase-matching and quasi-phase-matching, which assumes the equal interaction length of the nonlinear media and effective nonlinear coefficients. Under the phase-mismatching conditions, the amplitude of the SHG electric field will oscillate with a phase difference of $2\pi$ in one period. To address the oscillation of the electric amplitude, the phase-matching condition with $2k_1 = k_2$ can be achieved by the birefringence of the anisotropic crystals at an angle where the refractive indexes at the fundamental and SHG wavelengths are identical to each other $n_{2\omega}(z) = n_\omega(z)$, or the quasi-phase-matching can be achieved by adding the reciprocal vector ($G$) with periodic reversal polarization that corresponds the sign of $d_{eff}$ to satisfy $k_2 - 2k_1 - mG = 0$, where $m$ is an integer. Compared with the perfect phase-matching condition, the quasi-phase-matching has the obvious advantages in the use of the largest $d_{eff}$ in certain nonlinear crystals e.g. PPLN, PPLT and PPKTP[8,11].

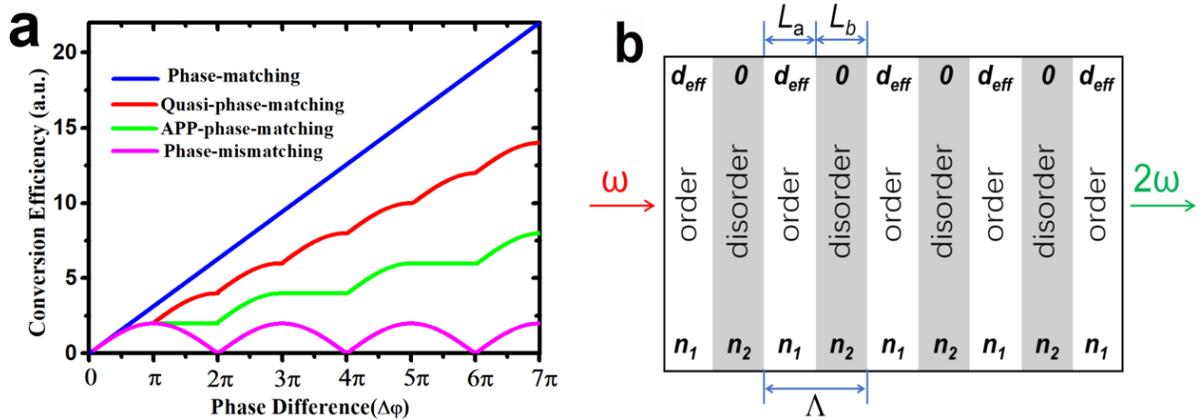

**Fig. 1 | Proposed APP phase-matching for the nonlinear frequency conversion. a**, Schematic demonstration of SHG performance under different phase-matching and phase-mismatching conditions assumed the equal interaction length of the nonlinear media and same efficient nonlinear coefficients $d_{eff}$. **b**,

Schematic graph of additional-phase-matching condition in arbitrary nonlinear optical crystals. The white and grey region represent ordered crystal and disordered amorphous, respectively.

In accordance with the theory of quasi-phase-matching, a condition with an additional periodic phase (APP) $\Delta\varphi_{APP}$ to the fundamental and SHG light can be proposed. The APP concept means that after the light propagating at the coherence length $l_c$, the generated phase difference $\Delta\varphi_{PD}$ was compensated by the additional phase difference $\Delta\varphi_{APP}$ with $\Delta\varphi_{APP}+\Delta\varphi_{PD}=0$ or 2mπ. The APP can be realized by periodically engineering regions in the nonlinear crystals to undermine the translational symmetry of the nonlinear crystal and block the conversion of the energy from the SHG to the fundamental light, as well as the oscillation of the SHG electric amplitude. The schematic demonstration for APP phase-matching is shown in **Fig. 1a,b**. In accordance with the phase-matching and quasi-phase-matching theory[2], only the APP with $\Delta\varphi_{APP}=(2m+1)\pi$ has a positive effect on the phase-matching. Similar to the transitional quasi-phase-matching but without the requirement of the reversal domains that only exist in some typical ferroelectric crystals, the proposed APP phase-matching exhibits remarkable advantages in relaxing the phase-matching requirements and using the largest efficient nonlinear coefficient $d_{eff}$.

Given the present condition of DUV coherent light, a crystal quartz was used to form the APP phase-matching. The $SiO_2$ crystal has the maximum nonlinear coefficient of $d_{11}$=0.3 pm/V[17], and its cutoff wavelength edge was measured below 150 nm with a 3 mm-length sample cut along with its Z direction. A 350 fs pulsed laser at 1040 nm was used for the laser writing, the amorphous $SiO_2$ would be produced in the broken regions[24] with suitable writing energy. An optical parametric oscillator (OPO, Opolette ™ HE 355 II) with the 10 ns pulsed laser at 355 nm served as the pump source. The APP quartz and power meter were placed into a chamber full of nitrogen to eliminate the absorption of oxygen. An additional $CaF_2$ prism was added to deflect and separate the fundamental (355 nm) and SHG (177.3 nm) signal (**Fig. 2a**). As aforementioned, the optical conversion efficiency strongly depends on the fluctuation of phase difference. Only $\Delta\varphi_{APP}=\Delta\varphi_{PD}=(2m+1)\pi$ is demanded to satisfy the phase-matching criteria (**Fig. 2b**). With the dispersion of the refractive index of $SiO_2$, the required period length $l_a$ (and $l_b$) for $\Delta\varphi_{APP} = \Delta\varphi_{PD} = \pi$ is 0.7 μm. However, this value has been smaller than minimum accuracy of femtosecond laser manufacturing. Therefore, an APP quartz with $l_a=l_b$=2.1 μm was utilized to generate 177.3 nm SHG laser, corresponding to phase difference $\Delta\varphi_{APP} = \Delta\varphi_{PD} = 3\pi$. Evidently, this conversion efficiency is lower than that case of phase difference of π. In order to prove the above theory, an APP quartz with $l_a=l_b$=1.4 μm was also utilized, corresponding to phase difference $\Delta\varphi_{APP} = \Delta\varphi_{PD} = 2\pi$.

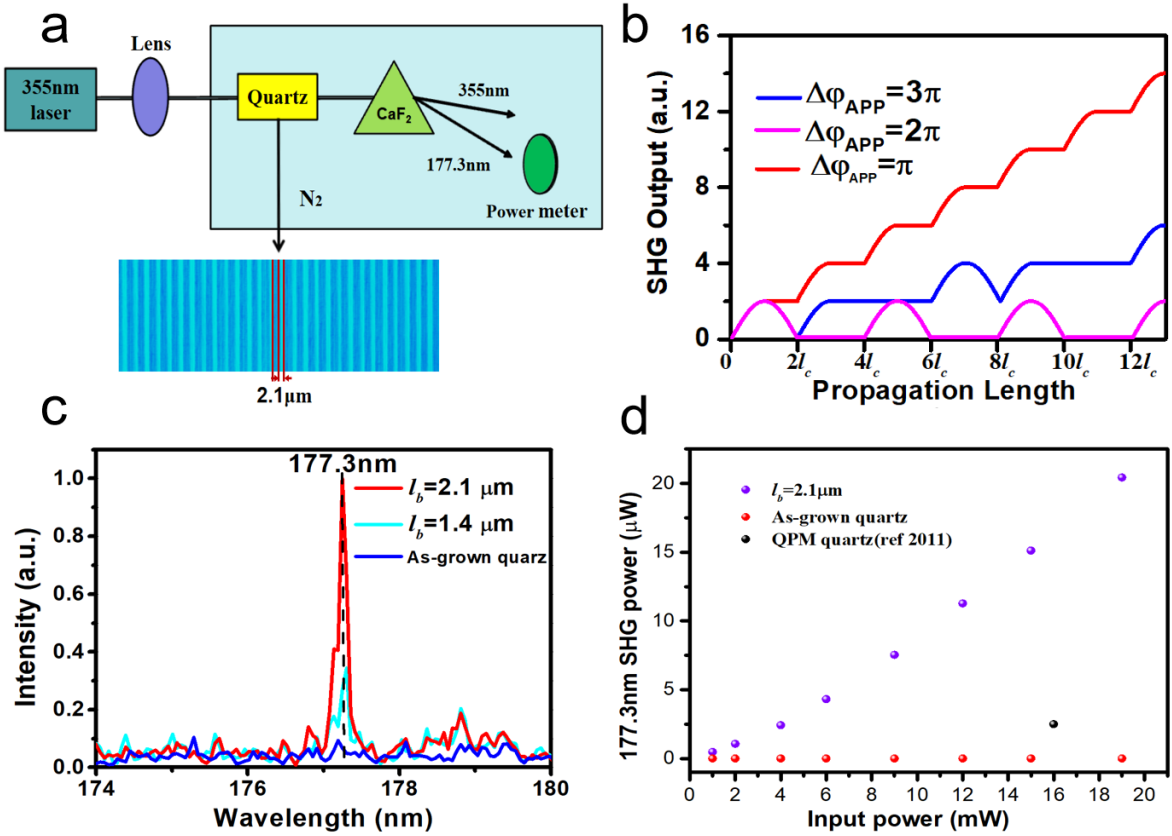

**Fig. 2 | SHG performance at the wavelength of 177.3 nm. a**, Experimental demonstration of APP SHG process in deep-ultraviolet region. **b**, Schematical estimation of the SHG output of the APP quartz with different shifted phase ($\Delta\varphi_{APP}$) under the same crystal length with coherent length $l_c$. **c**, SHG response signals (177.3 nm) in the APP quartz samples with APP $\Delta\varphi$ =3π ($l_b$=2.1 μm, red line) and $\Delta\varphi$ =2π ($l_b$=1.4 μm, cyan line) and as-grown quartz (blue line). **d**, 177.3 nm SHG output power in APP quartz (purple point) with $L_a=L_b$= 2.1 μm and $\Delta\varphi$=3π and as-grown quartz (red point). The 193 nm output in quasi-phase-matching twined quartz (black point) is adapted from ref[25], Copyright 2011 Optical Society of America.

As depicted in **Fig. 2c**, an anticipated peak locating at 177.3 nm emerges in sample quartz with APP Δφ =3π, which represents the first efficient DUV SHG output in quartz since 1961. In comparison, there is no any detectable SHG signal and very weak signal in as-grown crystalline quartz and sample quartz with APP Δφ =2π, suggesting our phase-compensating strategy is indeed significant. With the improvement of incident power $P_0$ up to 19 mW, the output SHG power $P_{2\omega}$ increases to 20.4 μW with the optical conversion efficiency $\eta$ of 1.07‰ at the maximum power (**Fig. 2d**). The output power and conversion efficiency could be further improved by improving the manufacturing process to facilitate the transmission and achieve the APP quartz with $\Delta\varphi_{APP} = \Delta\varphi_{PD} = \pi$. Even, this record efficiency is still seven times higher than that of 386 nm SHG in stressed twin quartz (0.153‰, black point in **Fig. 2d**)[25]. Accordingly, APP method is demonstrated to be a significant route to obtain deep-ultraviolet coherent lasers.

In summary, for the first time, a universal additional periodic phase (APP) for the phase-matching rule in the nonlinear optics was theoretically proposed, beyond the traditional phase-matching with natural birefringence of the crystals and quasi-phase-matching with reversible ferroelectric domains. APP technology is suitable for any acentric crystals (nonpolar, polar and ferroelectric phase) and amongst them, nonpolar nonlinear crystal is the best candidate to demonstrate this theory. In particular, the deep-ultraviolet 177.3 nm generation was firstly achieved via periodic disordered quartz crystal (nonpolar phase) with a high efficiency of 1.07‰. This APP strategy provides a versatile route for arbitrary nonlinear crystal in broadband wavelength. More importantly, this order/disorder alignment adds to a variable physical parameter into optical system, thus leading to next-generation revolution in nonlinear or linear modulation and classical or quantum photonics.

**References:**

1. Franken, P., Hill, A. E., Peters, C. W. & Weinreich, G. Generation of optical harmonics. *Phys. Rev. Lett.* **7,** 118-119 (1961).

2. Armstrong, J. A., Bloembergen, N., Ducuing, J. & Pershan, P. S. Interactions between light waves in a nonlinear dielectric. *Phys. Rev.* **127,** 1918-1939 (1962).

3. Maker, P. D., Terhune, R. W., Nisenoff, M. & Savage, C. M. Effects of dispersion and focusing on the production of optical harmonics. *Phys. Rev. Lett.* **8,** 21-22 (1962).

4. Boyd, R. W. Nonlinear Optics. (Elsevier, New York, 2003).

5. Nikitin, D. G., Byalkovskiy, O. A., Vershinin, O. I., Puyu, P. V. & Tyrtyshnyy, V. A. Sum frequency generation of UV laser radiation at 266 nm in LBO crystal. *Opt. Lett.* **41,** 1660-1663 (2016).

6. Petrov, V. Frequency down-conversion of solid-state laser sources to the mid-infrared spectral range using non-oxide nonlinear crystals. *Prog.Quant. Electron.* **42,** 1-106 (2015).

7. Shi, W. & Ding, Y. J. Continuously tunable and coherent terahertz radiation by means of phase-matched difference-frequency generation in zinc germanium phosphide. *Appl. Phys. Lett.* **83,** 848-850 (2003).

8. Zhu, S. N., Zhu, Y. Y. & Ming, N. B. Quasi-phase-matched third-harmonic generation in a quasi-periodic optical superlattice. *Science*, **278,** 843-846 (1997).

9. Wei, D. Z. et al. Experimental demonstration of a three dimensional lithium niobate nonlinear photonic crystal. *Nat. Photon.* **12,** 596-600 (2018).

10. Xu, T. X. et al. Three-dimensional nonlinear photonic crystal in ferroelectric barium calcium titanate. *Nat. Photon.* **12,** 591-595 (2018).

11. Petit, Y., Boulanger, B., Segonds, P. & Taira, T. Angular quasi-phase-matching. *Phys. Rev. A* **76,** 063817 (2007).

12. Yin, X. B. et al. Edge Nonlinear optics on a MoS2 atomic monolayer. *Science*, **344,** 488-490 (2014).

13. Pereira, S. F., Xiao, M., Kimble, H. & Hall, J. Generation of squeezed light by intracavity frequency doubling. *Phys. Rev. A* **38,** 4931-4934 (1988).

14. Kwiat, P. G. et al. New high-intensity source of polarization-entangled photon pairs. *Phys. Rev. Lett.* **75,** 4337-4341 (1995).

15. Zhang, X. Y. et al. Symmetry-breaking-induced nonlinear optics at a microcavity surface. *Nat. Photon.* **13,** 21-24 (2019).

16. Leuthold, J., Koos, C. & Freude, W. Nonlinear silicon photonics. *Nat. Photon*. **4,** 535-544 (2010).

17. Dmitriev, V. G., Gurzadyan, G. G. & Nikogosyan, D. N. Handbook of Nonlinear Optical Crystals (Springer, third Edition, 1999).

18. Suchowski, H. et al. Phase mismatch–free nonlinear propagation in optical zero-index materials. *Science,* **342,** 1223-1226 (2013).

19. Cyranoski, D. Materials science: China's crystal cache. *Nature*, **457,** 953-955 (2009).

20. Mei, L. et al. Crystal structure of $KBe_2BO_3F_2$. *Z. Kristallogr*. **210,** 93-95 (1995).

21. Meng, J. et al. Coexistence of Fermi arcs and Fermi pockets in a high-$T_c$ copper oxide superconductor, *Nature*, **462,** 335-338 (2009).

22. Xu, S. Y. et al. Observation of Fermi arc surface states in a topological metal. *Science,* **347,** 294-298 (2015).

23. Xu, B. et al. Generation of high power 200 mW laser radiation at 177.3 nm in $KBe_2BO_3F_2$ crystal. *Appl. Phys. B* **121,** 489-494 (2015).

24. Davis, K.M., Miura, K., Sugimoto, N. & Hirao, K. Writing waveguides in glass with a femtosecond laser. *Opt. Lett*. **21,** 1729-1731 (1996).

25. Kurimura S*, et al.* Quartz revisits nonlinear optics: twinned crystal for quasi-phase matching Invited. *Opt. Mater. Express* **1**, 1367-1375 (2011).